\documentclass{osa-article}

\journal{oe}

\usepackage{my-macro-equation}

\usepackage{bm}
\usepackage{graphicx} 

\newcommand\figref[1]{Fig.~\ref{fig:#1}}
\newcommand\figureref[1]{Figure~\ref{fig:#1}}
\newcommand\figuresref[1]{Figures~\ref{fig:#1}}

\usepackage{scalerel}
\usepackage{tikz}
\usetikzlibrary{svg.path}
\definecolor{orcidlogocol}{HTML}{A6CE39}
\tikzset{
  orcidlogo/.pic={
    \fill[orcidlogocol] svg{M256,128c0,70.7-57.3,128-128,128C57.3,256,0,198.7,0,128C0,57.3,57.3,0,128,0C198.7,0,256,57.3,256,128z};
    \fill[white] svg{M86.3,186.2H70.9V79.1h15.4v48.4V186.2z}
                 svg{M108.9,79.1h41.6c39.6,0,57,28.3,57,53.6c0,27.5-21.5,53.6-56.8,53.6h-41.8V79.1z M124.3,172.4h24.5c34.9,0,42.9-26.5,42.9-39.7c0-21.5-13.7-39.7-43.7-39.7h-23.7V172.4z}
                 svg{M88.7,56.8c0,5.5-4.5,10.1-10.1,10.1c-5.6,0-10.1-4.6-10.1-10.1c0-5.6,4.5-10.1,10.1-10.1C84.2,46.7,88.7,51.3,88.7,56.8z};
  }
}
\newcommand\orcidicon[1]{\href{https://orcid.org/#1}{\mbox{\scalerel*{
\begin{tikzpicture}[yscale=-1,transform shape]
\pic{orcidlogo};
\end{tikzpicture}
}{|}}}}

\usepackage{here}
\usepackage{hyperref} 
\begin{document}
\title{Characterization of entangling properties of quantum measurement via two-mode quantum detector tomography using coherent state probes}

\author{Shota Yokoyama,\authormark{1,2,*}\orcidicon{0000-0001-6708-3785}
Nicola Dalla Pozza,\authormark{1,2}\orcidicon{0000-0003-2764-9603}\\
Takahiro Serikawa,\authormark{1,3}
Katanya B.\ Kuntz,\authormark{1,2,4,5}\\
Trevor A.\ Wheatley,\authormark{1,2}\orcidicon{0000-0002-6878-2446}
Daoyi Dong,\authormark{1}\orcidicon{0000-0002-7425-3559}\\
Elanor H.\ Huntington,\authormark{1,2,6}\orcidicon{0000-0002-7154-0042}
and
Hidehiro Yonezawa\authormark{1,2,$\dagger$}\orcidicon{0000-0002-5522-1439}}

\address{\authormark{1}School of Engineering and Information Technology, The University of New South Wales, \\ Canberra, ACT 2600, Australia\\
\authormark{2}Centre for Quantum Computation and Communication Technology, Australian Research Council\\
\authormark{3}Department of Applied Physics, School of Engineering, The University of Tokyo, \\ 7-3-1 Hongo, Bunkyo-ku, Tokyo 113-8656, Japan\\
\authormark{4}Institute for Quantum Computing, University of Waterloo, Waterloo, Onatario N2L 3G1, Canada\\
\authormark{5}Department of Physics and Astronomy, University of Waterloo, Waterloo, Onatario N2L 3G1, Canada\\
\authormark{6}Research School of Electrical, Energy and Materials Engineering, College of Engineering and Computer Science, Australian National University, Canberra, ACT 2600, Australia}

\email{\authormark{*}s.yokoyama@unsw.edu.au} 
\email{\authormark{$\dagger$}h.yonezawa@unsw.edu.au} 
\homepage{https://www.hy-lab.net/} 

\begin{abstract}
Entangled measurement is a crucial tool in quantum technology.
We propose a new entanglement measure of multi-mode detection, which estimates the amount of entanglement that can be created in a measurement.
To illustrate the proposed measure, we perform quantum tomography of a two-mode detector that is comprised of two superconducting nanowire single photon detectors.
Our method utilizes coherent states as probe states, which can be easily prepared with accuracy. Our work shows that a separable state such as a coherent state is enough to characterize a potentially entangled detector.
We investigate the entangling capability of the detector in various settings.
Our proposed measure verifies that the detector makes an entangled measurement under certain conditions, and reveals the nature of the entangling properties of the detector. 
Since the precise characterization of a detector is essential for applications in quantum information technology, the experimental reconstruction of detector properties along with the proposed measure will be key features in future quantum information processing.
\end{abstract}

\section{Introduction}
Measurement device or technique is at the heart of rapidly developing quantum technologies, such as ultra-accurate sensing \cite{Giovannetti11}, absolutely secure or high-capacity communication \cite{Lo14,Chen12}, quantum computation \cite{Nielsen00}, and quantum simulation \cite{Georgescu14}.
While it is imperative to develop accurate and efficient measurement devices to observe quantum phenomena, measurements are also employed as part of the quantum process. 
There are several such examples; a linear optics quantum computation scheme that utilizes photon detections to achieve universal computation \cite{KLM01}, one-way quantum computation that employs a gigantic entangled state and successive measurements for computation \cite{Briegel01,Raussendorf01, Briegel09}, and the creation of non-classical states such as Schr\"{o}dinger cat states by projection onto number states  \cite{Ourjoumtsev06,Nielsen06,Wakui07}. 

Among the variety of quantum measurements in various settings, an {\it entangled measurement} (i.e., a projection onto an entangled basis) is a particularly important class of measurements. The Bell measurement is a typical example of an entangled measurement, which is a projective measurement on one of four Bell states (maximally entangled states of two qubits). 
The Bell measurement is one of the most fundamental building blocks for quantum information processing, and is highlighted in quantum teleportation \cite{Bennett93,Gottesman99}, entanglement swapping \cite{Zukowski93,Pan98}, quantum repeaters \cite{Briegel98, Duan01}, quantum key distribution \cite{Lo12}, and quantum parameter estimation \cite{Vidrighin14}. 
The entangling capability of a Bell measurement has a significant effect on the performances of these protocols.

Entangled measurements in the field of photonic quantum technology can be realized with a beam splitter network followed by single-mode detectors. By properly designing the beam splitter network and the number of detectors, various entangled measurements are implemented. This type of {\it assembled} entangled detector is well understood by analyzing the individual components and the overall structure.

On the other hand, there is a situation where we do not have any knowledge on the internal structure of a multi-mode measurement. Note that an actual detector normally has sensitivity over multiple modes including temporal, frequency, spatial, and polarization. Thus an actual detector may be considered as a multi-mode detector if we do not implement careful mode filtering to eliminate unwanted modes from entering the detector.

A possible interference between multiple modes inside a detector may lead to an entangled projection, typically ignored as a standard practice. 
Therefore, the detailed characterization of detectors, including a proper measure of entanglement, is of practical importance and of potential use in quantum information applications, as well as fundamental quantum physics. 

In this paper, we propose a new entanglement measure of a detector. This measure uncovers entanglement creatable by the act of measurement in a situation like entanglement swapping.
The proposed entanglement measure is intuitive and easy to calculate from Positive-Operator-Valued Measures (POVMs) of the detector \cite{Nielsen00,Furusawa11}.

We experimentally demonstrate two-mode quantum detector tomography by utilizing coherent states.
Quantum detector tomography typically utilizes well-qualified photon number states or coherent states, depending on the experimental system \cite{Luis99,Lundeen09}.
Since the experimental cost of preparing coherent states is less expensive than photon number states, 
coherent states are more suitable for 
quantum detector tomography involving huge parameters (e.g., $\sim 1.8\times 10^6$ in \cite{Zhang12}).
However, 
detector tomography using coherent states has not yet been demonstrated for entangled measurements.

In this paper, with coherent state probes, we characterize
a polarization two-mode detector consisting of two superconducting nanowire single photon detectors (SNSPDs), which emulates various two-mode detectors.
We reconstruct the POVMs of these detectors, and show under what conditions a detector can make an entangled measurement.
We note that Roccia et al. \cite{Roccia17} have demonstrated detector tomography of Bell measurements by utilizing qubits as input states, and estimated precisions of multiparameter estimations with the reconstructed measurements. 
In contrast to their work, our demonstration of the proposed entanglement measure focuses more on the general characterization of a detector, with implications in a wider range of quantum applications involving entanglement creation via measurements.

Our experimental results show that a completely mixed measurement of two orthogonal entangled measurements does not create entanglement. However, asymmetric measurements induced by photon loss may create entanglement. 
Furthermore, we characterize an SNSPD that has asymmetric polarization sensitivities.
These results, along with our developed detector tomography techniques and proposed
measure, give a deeper understanding and quantification of detector measurements, which
is useful in numerous complex quantum applications.

\section{Results}
\subsection{Proposed entanglement measure of a POVM}
Here we introduce our proposed entanglement measure. 
We assume a two-mode measurement where each mode has a $d$-dimensional Hilbert space. 
The measurement is fully described by a two-mode POVM $\{ \hat \Pi ^{(i)}_{A,B} \}$.
The probability of obtaining outcome $i$ is represented as $p_i = \Tr [ \hat \rho _{A,B}\hat \Pi ^{(i)}_{A,B} ]$,
where $\hat \rho_{A,B}$ is the density matrix of an input bipartite state consisting of modes $A$ and $B$.
In order to quantify the entangling capability of a detector, we define the entanglement measure of a POVM as,
\begin{align} 
	M (\hat \Pi ^{(i)}_{A,B})  \equiv 
	E \left( \frac{\hat \Pi ^{(i)}_{A,B}}{\mathrm{Tr} \left[ \hat \Pi ^{(i)}_{A,B} \right]}\right),
	\label{eq:Definition_of_M}
\end{align}
where $E$ represents an entanglement measure of a state. Note that a normalized POVM is mathematically the same as a density operator as it is a positive semidefinite and has a unit trace. Any entanglement measure of a state is applied to the normalised POVM if it is invariant under local unitary operations (see Appendix for details).
In this paper, the logarithmic negativity $E_{LN}$ is chosen because of the easiness of numerical calculation \cite{Vidal02,Plenio07}.

The logarithmic negativity of a quantum state is defined as 
$E_{LN} (\hat \rho _{A,B} ) = \log _2 \left\lVert \hat \rho_{A,B}^{T_B}\right\rVert$,
where $T_B$ is a partial transpose of mode $B$ \cite{Peres96},
and a trace norm is defined as $\lVert X \rVert \equiv \mathrm{Tr} \sqrt{X^\dagger X}$, which corresponds to the sum of absolute values of the eigenvalues of $X$.
The entanglement measure $M_{LN}$ is greater than or equal to zero. Only when the measurement is separable ($\hat \Pi_{A,B} = \hat \Pi_A \otimes \hat \Pi_B$), $M_{LN}$ reaches zero. Otherwise it indicates the {\it inseparability} of the measurement.

\subsection{Interpretation of proposed entanglement measure}
\begin{figure}[b]
	\centering
	\includegraphics[scale=1, clip]{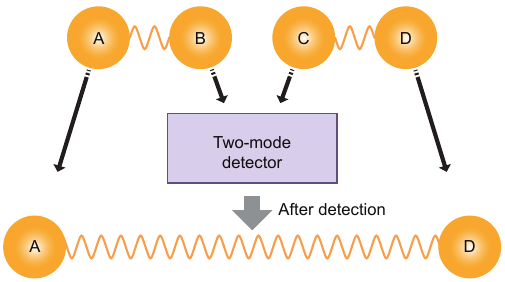}
	\caption{Two-mode measurement on two bipartite entangled states (i.e., entanglement swapping). Each circle and link between circle represent a quantum state and quantum entanglement, respectively. The inseparability of a two-mode detector is directly related to the maximum entanglement between modes $A$ and $D$, created by the measurement on modes $B$ and $C$.}
	\label{fig:ES}
\end{figure}
Next, we consider the meaning of this measure, and justify that the inseparability of the POVM is related to the entangling properties of the measurement. Let us apply a two-mode (or joint) measurement on two bipartite entangled pairs, as shown in \figref{ES}.
We assume that $A$-$B$ and $C$-$D$ are maximally entangled bipartite states. We perform a two-mode projective measurement on $B$ and $C$, and then consider the created entanglement between $A$ and $D$. After simple algebraic calculations, we find that the remaining state after measurement is proportional to the transposed POVM, up to local unitary operations (see Appendix).
Thus our proposed entanglement measure corresponds to entanglement of a state after the measurement under the condition that $E$ is invariant under local unitary operations, 
\begin{align}
	M (\hat \Pi ^{(i)}_{B,C}) 
	=
	E (\hat \rho ^{(i)}_{A,D}),
\label{eq:Mpi=Erho}
\end{align}
where $\hat \rho ^{(i)}_{A,D} = \mathrm{Tr}_{B,C}\left[ \hat \rho \hat \Pi_{B,C} ^{(i)}\right]/p_i$ is the remaining state after measurement,
$p_i = \Tr\left[\hat \rho \hat \Pi ^{(i)}_{B,C} \right]$ is the probability of obtaining an outcome $i$,
and the initial state $\hat \rho$ consists of two arbitrary maximally entangled pairs.
The inseparability of the POVM is directly related to the maximum entanglement created by the measurement.
It is worth noting that we consider conditionally created entanglement with the outcome $i$ of measurements. Although entanglement may be created conditionally by local operation and classical communications (LOCC), i.e., entanglement distillation \cite{Bennett96}, it is restricted to the case where modes $A$ and $D$ have initial (weak) entanglement. In our current scheme, LOCC cannot create entanglement between $A$ and $D$ even conditionally, because there is no initial entanglement. Thus the inseparability of the POVM, i.e., the created entanglement after the measurement, truly proves that the measurement is non-local and capable of creating entanglement.

We note again that our entanglement measure is defined for each POVM element. Entanglement creatable by the measurement may differ for each POVM element, i.e., the  entanglement depends on the measurement outcome $i$. In the sense of characterizing the entangling capability of the detector (not each POVM element), it is also possible to  consider the averaged entanglement measure of a POVM (e.g., $\sum_i p_i M (\hat \Pi ^{(i)})$) by assuming the particular set of inputs.

\subsection{Experiment: Two-mode quantum detector tomography with coherent states}
\begin{figure}[b]
	\centering
	\includegraphics[scale=1.12, clip]{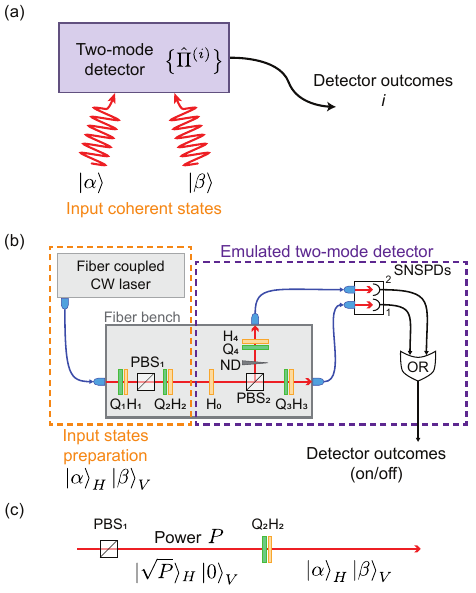}
	\caption{
		Characterization of multi-mode detector. (a) Abstract illustration of two-mode quantum detector tomography.
		Input states are two-mode coherent states $\ket{\alpha}\ket{\beta}$. $\alpha$ and $\beta$ are complex amplitudes of each coherent state.
		(b) Experimental setup. 
		(c) Details of input states preparation.
		Input coherent states are horizontally and vertically polarization two-modes $\ket{\alpha}_H\ket{\beta}_V$ within an optical beam.
			PBS, Polarization Beam Splitter;
			H, Half Wave Plate;
			Q, Quarter Wave Plate;
			ND, Neutral Density Filter;
			SNSPD, Superconducting Nanowire Single Photon Detector.
			Optical beams are transmitted through single-mode optical fibers except inside a fiber bench.
	}
	\label{fig:ExperimentalSetup}
\end{figure}
In order to demonstrate the usefulness of our entanglement measure of a POVM, we experimentally construct various two-mode quantum detectors. We then reconstruct the POVMs via quantum detector tomography utilizing coherent states as inputs.
We stress that our aim is not to generate entangled states by the measurements, or to characterize the entangled \textit{state}. Our goal is to characterize the multi-mode \textit{detector} and reveal its entangling capability.
Therefore, our focus is on investigating what kind of measurement has the ability of creating entanglement.

\figureref{ExperimentalSetup}(a) shows an abstract illustration of two-mode quantum detector tomography.
POVMs are reconstructed via the relationship of $p_i = \Tr [ \hat \rho \hat \Pi ^{(i)} ]$, where the probability $p_i$ is experimentally obtained and $\hat \rho$ is the known input states (i.e., this is an inverse problem).
Coherent states provided by a laser may be a good input probe because they can be easily prepared with well-defined amplitudes and phases.
Interestingly, a separable state such as a product of coherent states is enough to characterize a possibly entangled detector.
However, since we are reconstructing the POVM coefficients in the number state basis, care must be taken into account when truncating the number state representation of the coherent states, resulting in a limitation of the available amplitudes.

We record detector outcomes for various input states, and then solve a convex optimization problem to obtain the POVMs  \cite{Lundeen09,Roccia17,Zhang12f,Zhang12,Natarajan13,Brida12,Humphreys15,Feito09,Ansari17}. 
The amount of input (coherent) states depends on the degrees of freedom of the POVMs, and
additional constraints reduce the complexity of the reconstruction.
For simplicity, we assume that a two-mode detector has no phase reference for inputs (i.e., global phase insensitive detector), but it may have the sensitivity for the relative phase $\delta$ between two-mode inputs.
In this case, we can omit the global phase of two-mode input (coherent) states.
The general two-mode coherent state without global phase is expressed as $\ket{|\alpha|, |\beta| \ee^{i \delta}}\  (\delta \in \mathbb{R})$, which is expanded in the photon number basis as,
\begin{align}
\ket{|\alpha|, |\beta| \ee^{i \delta}} = 
	      \exp \left[  -\frac{1}{2} \left( |\alpha|^2 +|\beta|^2 \right)
  	          \right]
  	          \sum_{m,n}^{\infty} \frac{|\alpha|^m |\beta|^n \ee^{i n \delta}}
  	                                   {\sqrt{m! n!}} 
  	                              \left| m,n \right\rangle.
\label{eq:2modesCoherentState}
 \end{align}
The minimal set of inputs can be derived from the analytical expression of the POVM (see Appendix).
We also assume that each input mode has a two-dimensional Hilbert space ($d=2$), and the POVM is represented with a 4 by 4 matrix,
$\Pi^{(i)} = \Sigma \ \pi_{k,l,m,n}^{(i)} \crossprojection{k,l}{m,n}$ $(k,l,m,n =$ 0 or 1),
which is the minimum dimension to calculate our entanglement measure of a POVM.
However, we first reconstruct a 6 by 6 matrix representation, and then truncate it to a 4 by 4 matrix. 
This is because an input coherent state with a non-negligible amount of $\left| 1,1 \right>$ element  has $\left| 0,2 \right>$ or $\left| 2,0 \right>$ elements with the same order of magnitude  due to the Poisson photon statistics of a coherent state. 
Note that the POVM with a 6 by 6 matrix representation is not directly used for quantifying entanglement because it is not decomposed into two subsystems.

\subsection{Experimental setup of two-mode detector tomography}
\figureref{ExperimentalSetup}(b) shows our experimental setup.
We employ a continuous-wave (CW) fiber coupled laser at 1548.56~nm as the primary light source. 
The input state is a weak coherent state with an average photon number much less than one. 
The yellow dashed box in \figref{ExperimentalSetup}(b) and \figref{ExperimentalSetup}(c) correspond to the preparation of input states. 
After the polarization beam splitter (PBS$_1$), the horizontally polarized beam with power $P$, described as $|\sqrt{P}\rangle_H \ket{0}_V$, is converted into $\ket{|\alpha|}_H \ket{|\beta| \ee^{i\delta}}_V$ via a quarter wave plate (QWP$_2$) and a half wave plate (HWP$_2$), denoted as Q$_2$ and H$_2$ respectively.
We use 19 sets of coherent states for the inputs, which are represented by the parameters in the set,
\begin{align}
\notag
&\{(|\alpha| ,|\beta|, \delta) \} = 
\Bigg\{(0,0,-), 
\left(\sqrt{\tfrac{P_1}{2}},\sqrt{\tfrac{P_1}{2}}, \tfrac{m_1\pi}{4}\right),
\left(\sqrt{P_1},0,- \right),
\left(0,\sqrt{P_1},-\right),\\
& \qquad 
\left(\sqrt{\tfrac{P_2}{4}},\sqrt{\tfrac{3P_2}{4}}, \tfrac{m_2\pi}{2}\right),
\left(\sqrt{\tfrac{3P_2}{4}},\sqrt{\tfrac{P_2}{4}}, \tfrac{m_2\pi}{2}\right)
\biggr | m_1=\{ 0,1,\cdots, 7 \}, m_2=\{ 0,1,2, 3 \}
\Bigg\},\label{eq:set}
\end{align}
where we use a vacuum input and two powers, $P_1 = 0.20$~photons and $P_2=0.05$~photons.
Note that we decide to employ a slightly larger number of coherent states than the minimum number of 14, 
which corresponds to the degrees of freedom of POVMs with a 6 by 6 matrix representation under the constraint of global phase insensitivity $\pi_{k,l,m,n}^{(i)} = 0$ ($k+l \neq m+n$),
to simplify the rotating angles of two wave plates (see Table \ref{tb:WavePlateSet} in Appendix).

The dashed purple box in \figref{ExperimentalSetup}(b) corresponds to the two-mode quantum detector 
comprising of two SNSPDs ({\it Photon Spot, Inc.}), PBS$_2$, multiple wave plates and a neutral density (ND) filter.
Our detector emulates various two-mode measurements by introducing loss in one channel.  The polarization of the input beam is first rotated by HWP$_0$ with the azimuth angle of $22.5^\circ$, which changes, for example, a horizontally polarized beam into a diagonally polarized beam.
PBS$_2$ following HWP$_0$ splits the optical beam into two spatially separated beams. Various ND filters with corresponding loss $L$ are inserted in one of the two beams.   
Two optical beams are then injected into two SNSPDs through single-mode optical fibers. 
Note that we maximize the quantum efficiencies of the SNSPDs by rotating QWP$_{3,4}$ and HWP$_{3,4}$ placed after PBS$_2$. 
The photon counting signals from the two SNSPDs are then sent to a logical OR gate, and the final detector output is obtained as an on/off signal.

\begin{figure}[t]
	\centering
	\includegraphics[scale=1.12, clip]{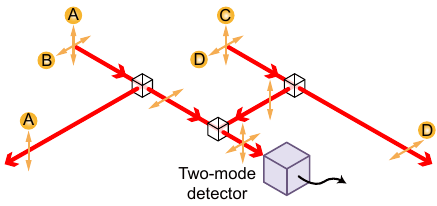}
	\caption{
		Example of an entanglement swapping setup by single-rail encoding.
		Mode $A$ ($C$) and mode $B$ ($D$) are encoded as horizontally and vertically polarized beams, respectively. 
		Our two-mode quantum detector described in \figref{ExperimentalSetup}(b) can be used as a projective measurement on modes $B$ and $C$.
	}
	\label{fig:SingleRailES}
\end{figure}

This detector has two extreme settings, i.e., separable measurement or maximally entangled measurement.
First let us assume that we do not have an ND filter in the beam ($L=0$). 
If the input beam has at most one photon, the photon detection at SNSPD$_{1(2)}$ is actually a projection onto the single-rail encoded Bell state $\ket{\Psi^+}$ ($\ket{\Psi^-}$) up to one photon elements $\pi^{(i)}_{k,l,m,n} (\max (k+l,m+n)\le 1)$, where 
  $\ket{\Psi^\pm} = (\ket{1}_H\ket{0}_V\pm\ket{0}_H\ket{1}_V)/\sqrt{2}$. 
For example, this detector can be used as a projective measurement in an entanglement swapping experiment, as shown in \figref{SingleRailES}.
We note that the higher photon number POVM elements would differ from a projection onto the Bell state due to the lack of photon number resolution.
After detection by the SNSPDs, we apply a logical OR operation to the outcome signals. Therefore, the detector outcomes do not tell us which SNSPD detected a photon, as long as the two SNSPDs have the same quantum efficiency.
Therefore, the detector is a fully separable measurement mixing two projections onto $\ket{\Psi^+}$ and $\ket{\Psi^-}$.
By stark contrast, the detector becomes a maximally entangled measurement when we block one of the beams (i.e., $L=1$). This corresponds to the Bell measurement projecting onto either $\ket{\Psi^+}$ or $\ket{\Psi^-}$ up to one photon elements.
In addition, by changing the ND filter ($0<L<1$), we can continuously realize an in-between setting for the measurement type.
Note, again, that our maximally entangled measurement is not exactly the same as a true Bell measurement due to the lack of photon number resolution.

Our overall detector, i.e., the outcome after the logical OR operation, is an on/off detector, so it only has two POVMs, $\hat \Pi^{({\rm on})}$ and $\hat \Pi^{{\rm (off)}}$. We reconstruct one POVM, $\hat \Pi ^{(\mathrm{on})}$, under the constraint $0\le \hat \Pi ^{(\mathrm{on})}\le \hat 1$, and then calculate the other POVM, $\hat \Pi ^{(\mathrm{off})}=\hat 1 - \hat \Pi ^{(\mathrm{on})}$, so that the completeness relation of the POVMs is automatically satisfied.

In our setup, we use a CW laser rather than a pulsed laser, which was used in previous detector tomography experiments \cite{Zhang12f, Lundeen09,Zhang12,Natarajan13,Brida12,Humphreys15,Feito09,Ansari17}.
Compared with pulsed laser experiments, continuous wave lasers have an additional degree of freedom, i.e., we can arbitrarily determine a temporal mode of a quantum state. Since the detector characteristics are dependent on the temporal mode of an input state, this degree of freedom gives us an additional capability of characterization of a detector. 
In our experiment, we set the temporal mode as a square shape with the width of 1~$\mu$s for simplicity.
Note that the dead time of SNSPDs ($\sim$40~ns) is negligible since it is much shorter than the width of the wave packets.
The dark counts of our detectors are below 200 counts per second. Since 
we choose 1 $\mu$s wave packet, the dark count per wave packet is less than $2\times 10^{-4}$. Thus, the effect of the dark count is negligible in the current experiment.

We record 100,000 measurement outcomes with a 1~$\mu$s time window for each input state (1,900,000 outcomes in total) to reconstruct the POVMs.
We repeat each measurement run six times to derive experimental error bars (standard deviations) of the reconstructed POVMs.

\subsection{Reconstructed POVMs}
\begin{figure*}[t]
	\centering
	\includegraphics[scale=1.0,clip]{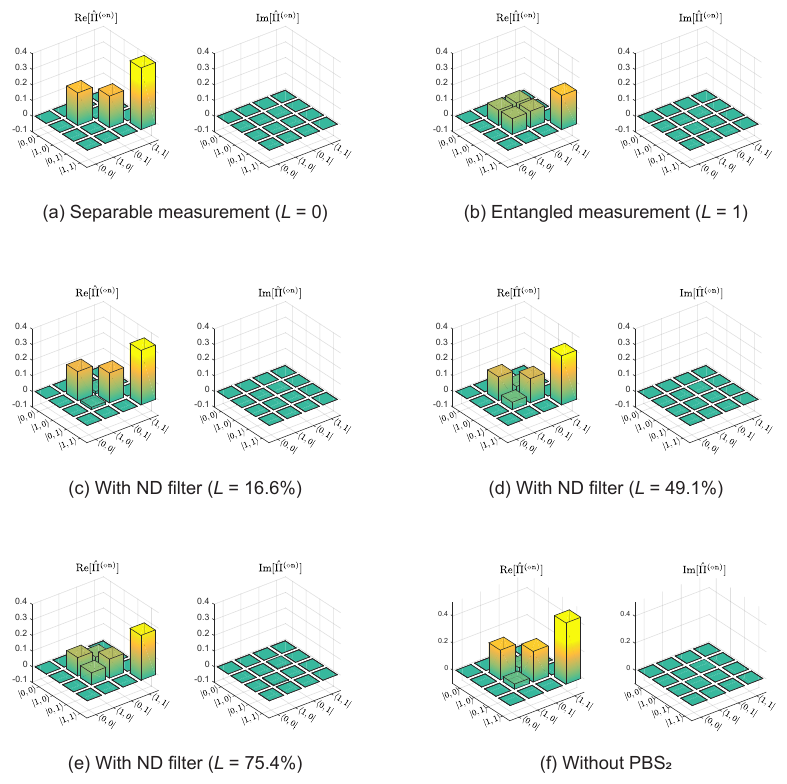}
	\caption{Reconstructed two-mode detector POVMs corresponding to ``on''.
			(a) Separable measurement with $L=0$ (no ND filter).
			(b) Entangled measurement with $L=1$ (blocking the reflected beam from PBS$_2$).
			(c)-(e) Other measurements with ND filters $L=16.6\%, 49.1\%, 75.4\%$.
			(f) Measurement without PBS$_2$.
			We omit the subscriptions denoting polarization modes for simplicity, $\ket{k,l} \equiv \ket{k}_H\ket{l}_V$ and $\bra{m,n} \equiv \mbox{}_H\! \bra{m}\mbox{}_V\!\bra{n}$.}
	\label{fig:POVM}
\end{figure*}

\figureref{POVM}(a) shows the reconstructed POVM $\hat \Pi ^{(\mathrm{on})}$ without an ND filter ($L=0$), which corresponds to a fully separable measurement. 
In the following, we omit the subscriptions denoting polarization modes for simplicity, $\ket{k,l} \equiv \ket{k}_H\ket{l}_V$ and $\bra{m,n}\equiv\mbox{}_H\! \bra{m}\mbox{}_V\!\bra{n}$.
This measurement can be considered as a mixture of projections onto two Bell states, $\frac{1}{2} \projection{\Psi^+} + \frac{1}{2} \projection{\Psi^-} = \projection{1,0}+\projection{0,1}$, if the input beam has up to one photon per wavepacket.
The diagonal elements of $\projection{1,0}$ and $\projection{0,1}$ are $0.209\pm 0.007$, and $0.201 \pm 0.008$, respectively, which correspond to the quantum efficiencies of the two SNSPDs. Those values agree with the theoretical predictions of $20\pm 1 \% $, which includes optical losses such as coupling inefficiency at the fiber coupler; the quantum efficiencies of the SNSPDs themselves are around $30\%$. We obtain 99.9 $\pm$ 0.05\% fidelity between the experimental and theoretical POVMs, which are derived by including the SNSPDs' finite quantum efficiencies and optical losses of the detector.
This verifies the high accuracy of our reconstruction. 

\figureref{POVM}(b) shows the reconstructed POVM of the entangled measurement.
In this case, we introduce 100\% loss ($L=1$) at one SNSPD. 
This measurement can be considered as a projection onto the Bell state, $\projection{\Psi^+} = (\projection{1,0}+\crossprojection{1,0}{0,1}+\crossprojection{0,1}{1,0}+\projection{0,1})/2$, if the input beam has up to one photon per wavepacket.
The reconstructed POVM has non-zero off-diagonal elements in contrast to the POVM of the separable measurement in
\figref{POVM}(a), which is evidence of the measurement's entangling properties.
The calculated fidelity is 99.5 $\pm$ 0.5\%.

\figuresref{POVM}(c)-\ref{fig:POVM}(e) show the reconstructed POVM of other measurements with ND filters.
Corresponding losses are \figref{POVM}(c) 16.6\%, \ref{fig:POVM}(d) 49.1\%, and \ref{fig:POVM}(e) 75.4\%.
We also perform experiments by removing PBS$_2$ from the setup shown in \figref{ExperimentalSetup}(b).
\figureref{POVM}(f) is the result without PBS$_2$.
In this configuration, the detector consists of only a single SNSPD$_1$ and does not have an artificial polarization-selective component. Thus, the reconstructed POVM represents the characteristics of the SNSPD itself.

\subsection{Photon detection probability for a maximally mixed state}
From the reconstructed POVMs, we can calculate the photon detection probabilities (i.e., the success probability of creating entanglement) for a particular input. As an example, we assume that the input state is fully randomized among $\ket{0,0}$, $\ket{1,0}$, $\ket{0,1}$, and $\ket{1,1}$, that is, a maximally mixed state $\sum_{k,l \in \{0,1\}}\projection{k,l}/4$. In this case, the photon detection probability is calculated from a trace of the POVM normalized by the dimension.

\begin{figure}[t]
	\centering
	\includegraphics[scale=0.6,clip]{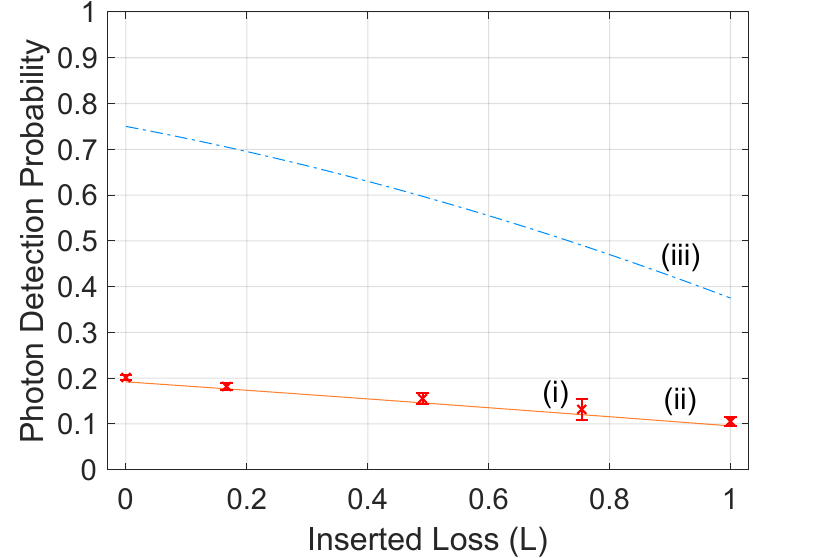}
	\caption{Photon detection probabilities of the reconstructed POVMs for a maximally mixed state.
			(i) Experimental results for various channel losses.
			Theoretical prediction curves (ii) with actual quantum efficiencies and (iii) ideal quantum efficiencies.
	}
	\label{fig:Prob}
\end{figure}

\figureref{Prob} shows the calculated photon detection probabilities of reconstructed POVMs as a function of the inserted loss $L$. The red crosses (i) show the experimental results, which show good agreements with the theory curve (ii) derived by taking into account the finite quantum efficiencies. The blue dash-dotted line (iii) is the theoretical curve assuming 100\% quantum efficiencies. At zero loss $L=0$, the photon detection probability is $0.75$, which means that the photon can be detected except for the case of the vacuum input. As the inserted loss increases, the probability decreases steadily, and it finally reaches $3/8=0.375$ at $L=1$. Note that the photon detection probability without PBS$_2$ is $0.225\pm 0.005$.

\subsection{Entanglement measure of reconstructed POVMs}
\begin{figure}[t]
	\centering
	\includegraphics[scale=0.6,clip]{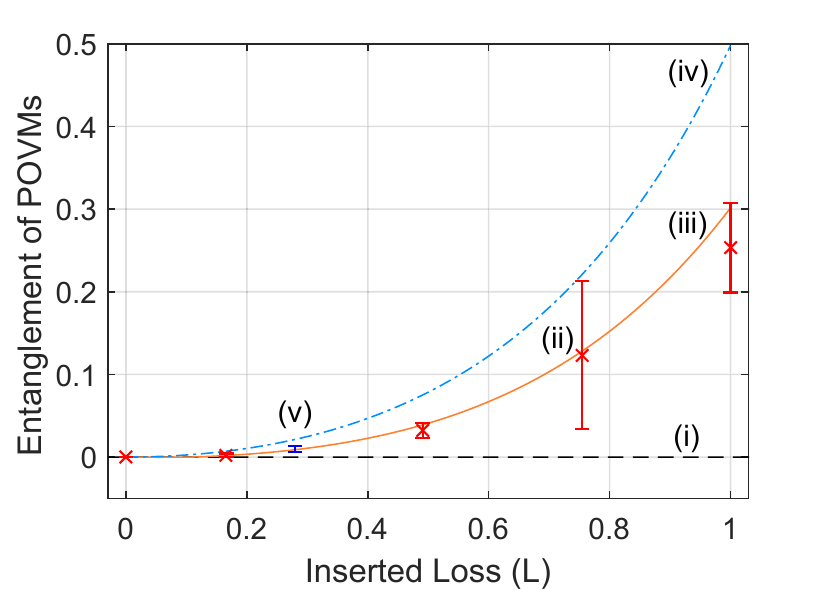}
	\caption{Logarithmic negativities of the POVMs.
			(i) Bound for non-entangled measurement.
			(ii) Experimental results for various channel losses.
			Theoretical prediction curves (iii) with actual quantum efficiencies and (iv) ideal quantum efficiencies.
			(v) Experimental result without PBS$_2$. 
	}
	\label{fig:LN}
\end{figure}

\figureref{LN} shows the entanglement measure $M_{LN} (\hat \Pi ^{(\mathrm{on})}) $ as a function of the inserted loss $L$.
Trace (i) shows the lower bound of zero, which indicates a separable or non-entangled measurement.
The red crosses (ii) show the experimental results. At zero loss $L=0$, $M_{LN}$ is zero, meaning a fully separable measurement. By increasing losses $L$ in one channel, $M_{LN}$ increases until it reaches a maximum of $0.25 \pm 0.05$ at $L=1$. This behavior may be counterintuitive because losses generally degrade entanglement. This result, however, indicates that some form of asymmetry is necessary to create entanglement. In the current setting, the selectivity of two orthogonal projective measurements ($\ket{\Psi^+}$ and $\ket{\Psi^-}$) is key to generate the entangled projection, and it is achieved by introducing loss in one channel.

The experimental results are well within error of the theory curve (iii), which is derived by taking into account the finite quantum efficiencies.
The blue dash-dotted line (iv) is the theoretical curve assuming 100\% quantum efficiencies. 
Note that this measurement cannot reach unity even with 100\% detection efficiencies, which can only be obtained by a true Bell measurement. This is due to the $\projection{1,1}$ element of the POVMs that are non-zero for our detector, which would be zero for a true Bell measurement. This non-zero $\projection{1,1}$ element is attributed to the lack of photon number resolution by the SNSPDs. 

The measurement result without PBS$_2$ is $M_{LN}=0.009 \pm 0.004$.
The non-zero $M_{LN}$ implies that an SNSPD itself could be an entangled measurement. This entangling capability stems from the asymmetric quantum efficiencies of two polarization modes in an SNSPD. This is the same situation as two SNSPDs with a finite loss $L$ in the previous results.
The ratio of the quantum efficiencies for orthogonal polarizations is measured as 0.721, corresponding to $L=0.279$. The red data point (v) shows the experimental result plotted at $L=0.279$, which is on the theoretical curve of (iii). 
We note that, in addition to the asymmetry of the polarization sensitivities, the polarization mixing at HWP$_0$ is essential for the non-zero $M_{LN}$. The entangled measurement is attributed to multi-mode interference and the choice of a particular projection basis.

\section{Conclusion}
We proposed an entanglement measure of a detector, and applied it to a two-mode detector. Our newly proposed measure represents the detector's ability to create entanglement. The experimental results imply that the entangling properties of the measurement stem from an asymmetry inside the detector. Our detector tomography technique and proposed measure can be extended to more than two modes, as well as to other physical modes such as frequency and temporal.
Precise characterization of the measurement detector employed is essential for applications in quantum information processing. These experimental demonstrations, along with our proposed measure, will be useful in future quantum information technologies.

\appendix
\section*{Appendix}
\section{Entanglement created by measurement}
We will explain how entanglement created by measurement is related to its POVM. As explained in the Results section, 
we assume that we have two pairs of maximally entangled states. We apply a two-mode measurement on them, and then examine the entanglement of the post-measurement state.

We assume $d$-dimensional Hilbert space for each mode. 
Arbitrary maximally entangled bipartite states are expressed with a particular maximally entangled state and local unitary operators as,
   \begin{align}
     \ket{\psi_M} &=\left( \hat U_A \otimes \hat V_B \right) \ket{\Phi ^+}_{A,B} \\
                  &=
     \frac{1}{\sqrt{d}}\sum_{\mu,\nu,n}u_{\mu,n}v_{\nu,n}\ket{\mu,\nu}_{A,B},
    \end{align}
where  $\ket{\Phi^+} = \tfrac{1}{\sqrt{d}}\sum_n\ket{n,n}$, and $\hat U_A $ ($\hat V_B$) is a local unitary operator for mode $A$ ($B$) \cite{Furusawa11}. 
$u_{\mu,n}$ and $v_{\nu,n}$ are the coefficients describing these unitary operators, i.e.,
 $\hat U \ket{n} = \sum_{\mu} u_{\mu,n}\ket{\mu}$ and $\hat V \ket{n} = \sum_{\nu} v_{\nu,n}\ket{\nu}$.
 
We have two pairs of maximally entangled states $\ket{\psi_1}$ and $\ket{\psi_2}$ as,
   \begin{align}
     \hat \rho &=\projection{\psi_1}_{A,B} \otimes \projection{\psi_2}_{C,D}, \\
     \ket{\psi_1}_{A,B} & = \frac{1}{\sqrt{d}}\left( \hat U^{(1)}_A \otimes \hat V^{(1)}_B  \right)\sum_{n_1}\ket{n_1,n_1}_{A,B}, \\
     \ket{\psi_2}_{C,D} & = \frac{1}{\sqrt{d}}\left( \hat U^{(2)}_C \otimes \hat V^{(2)}_D \right)\sum_{n_2}\ket{n_2,n_2}_{C,D},
   \end{align}
where $\hat \rho$ is the initial four-mode state. We apply a joint measurement on modes $B$ and $C$. The POVM of the joint measurement is expressed as,
   \begin{align}
     \hat \Pi_{B,C}= \sum_{i,j,k,l} \pi_{i,j,k,l} \crossprojection{i,j}{k,l}_{B,C}.
    \label{eq:POVMBC}
   \end{align}
Here we omit the superscript denoting the measurement outcome for simplicity.

The remaining state after the measurement is calculated as,
\begin{align}
\notag
	&
\Tr _{B,C}\left[\hat \rho\, \hat \Pi_{B,C}  \right]
= 
	\sum_{s,t} \bra{s,t}_{B,C}  \hat \rho\, \hat \Pi_{B,C} \ket{s,t}_{B,C}\\
\notag
	&\quad = 
	\sum_{s,t,i,j,k,l} \bra{s,t}_{B,C}  \hat \rho\, \pi_{i,j,k,l}  \ket{i,j}_{B,C} \bracket{k,l}{s,t}\\
\notag
	&\quad = 
	\sum_{i,j,k,l} \bra{k,l}_{B,C}  \hat \rho\, \pi_{i,j,k,l}  \ket{i,j}_{B,C} \\
\notag
	&\quad=
	\frac{1}{d^2}\sum_{i,j,k,l,n_1,n_2,n'_1,n'_2} \pi _{i,j,k,l}\left(\hat U^{(1)}_A \otimes \hat V^{(2)}_D  \right)
	\left[ \bra{k,l}_{B,C}\left(\hat V^{(1)}_B \otimes \hat U^{(2)}_C \right)
	\ket{n_1,n_2}_{B,C}\right]\\
	&
\qquad
	\times 
	\left[ \bra{n'_1,n'_2}_{B,C}\left(\hat V^{(1)}_B \otimes \hat U^{(2)}_C \right)^\dagger
	\ket{i,j}_{B,C}\right]
\label{eq:RemainingState}
	\left[ \crossprojection{n_1,n_2}{n'_1,n'_2}_{A,D}
	\left(\hat U^{(1)}_A \otimes \hat V^{(2)}_D  \right)^\dagger \right],
\end{align}
where $\Tr _{B,C}$ is a partial trace regarding modes $B$ and $C$. Note that this expression is not normalized.
Let us first calculate the part of Eq.~\eqref{eq:RemainingState},
\begin{align}
\notag
	\sum_{n_1,n_2} \bra{k,l}_{B,C}\left(\hat V^{(1)}_B \otimes \hat U^{(2)}_C \right)
	\ket{n_1,n_2}_{B,C} \ket{n_1,n_2}_{A,D}
	&
	=\sum_{n_1,n_2}  v^{(1)}_{k,n_1}u^{(2)}_{l,n_2}\ket{n_1,n_2}_{A,D} \\
	&
	=\left( \hat V'_{A} \otimes \hat U'_{D}\right) \ket{k,l}_{A,D},
	\label{eq:RemSt2nd}
\end{align}
where we define the new unitary operators as $\hat U'_D \ket{n} = \sum_{\mu} u^{(2)}_{n,\mu}\ket{\mu}$ and $\hat V'_A \ket{n} = \sum_{\nu} v^{(1)}_{n,\nu}\ket{\nu}$. 
In the same manner, the other part of Eq.~\eqref{eq:RemainingState} is calculated as,
\begin{align}
	&\sum_{n'_1,n'_2} \bra{n'_1,n'_2}_{A,D} \bra{n'_1,n'_2}_{B,C}\left(\hat V^{(1)}_B \otimes \hat U^{(2)}_C \right)^\dagger
	\ket{i,j}_{B,C}
	=\bra{i,j}_{A,D} \left( \hat V'_{A} \otimes \hat U'_{D}\right)^\dagger.
	\label{eq:RemSt3rd}
\end{align}
By substituting Eqs.~(\ref{eq:RemSt2nd}) and (\ref{eq:RemSt3rd}) into Eq.~\eqref{eq:RemainingState}, we obtain, 
  \begin{align}
	\notag
    \qquad \Tr _{B,C} \left[\hat \rho\, \hat \Pi_{B,C}  \right]  &=  
	\frac{1}{d^2} \sum_{i,j,k,l} \pi_{i,j,k,l}  \left(\hat U^{(1)}_A\hat V'_{A} \otimes \hat V^{(2)}_D\hat U'_{D} \right)\\
	& \qquad \qquad 
	\times 
                      \crossprojection{k,l}{i,j}_{A,D}
                      \left(\hat U^{(1)}_A\hat V'_{A} \otimes \hat V^{(2)}_D\hat U'_{D}\right)^\dagger, \notag \\
   &
	= \mathcal{U} \left(\hat \Pi_{A,D}\right) ^T \mathcal{U} ^\dagger,
 \end{align}
where we define the POVM $\hat \Pi_{A,D}$ and the local unitary operator $\mathcal{U}$ as,  
   \begin{align}
      \hat \Pi_{A,D} & = \sum_{i,j,k,l}  
                                      \pi_{i,j,k,l}\crossprojection{i,j}{k,l}_{A,D}, \\
      \mathcal{U} & \equiv 
                        \hat U^{(1)}_A\hat V'_{A} \otimes \hat V^{(2)}_D\hat U'_{D}.
   \end{align}
Note that this POVM $\hat \Pi_{A,D}$ is the same as the original POVM $\hat \Pi_{B,C}$ [Eq.~\eqref{eq:POVMBC}] except for the mode subscripts.
Finally we obtain the normalized density matrix after the measurement as, 
\begin{align}
	\hat \rho_{A,D} &=\frac{\Tr _{B,C}\left[\hat \rho\, \hat \Pi_{B,C}  \right] }{\Tr \left[\hat \rho\, \hat \Pi_{B,C}  \right] } =
\frac{\mathcal{U} \left(\hat \Pi_{A,D}\right) ^T \mathcal{U} ^\dagger }{\Tr \left[\hat \Pi_{A,D}\right]}. 
\end{align}
Since $\mathcal{U}$ is the local unitary operator, any entanglement measure $E (\hat \rho _{A,D}) $ that is invariant under the local unitary operations is directly applicable to the normalized POVM $\hat \Pi_{B,C}$ to quantify the entanglement created by the measurement. 
We obtain the final expression as,
\begin{align} 
	\notag
	E (\hat \rho _{A,D}) &= E \left( 
	     \frac{\mathcal{U} \left(\hat \Pi_{A,D}\right) ^T \mathcal{U} ^\dagger }{\Tr \left[\hat \Pi_{A,D}\right]}
	                           \right) \\
	&= E \left( 
	    \frac{ \hat \Pi_{B,C} }{\Tr \left[\hat \Pi_{B,C}\right]}
	     \right) \notag \\
	&=M(\hat \Pi_{B,C}).
\end{align}
Here $M$ is the entanglement measure of the POVM as defined in Eq.~\eqref{eq:Definition_of_M}. We relabelled the POVM $\hat \Pi_{A,D} $ as $\hat \Pi_{B,C}$, and note that $E$ is invariant for the transpose. 

\section{Analytical solution of POVM with coherent state inputs}
We will discuss how we can {\it analytically} calculate the POVM from the measurement probabilities with various coherent state inputs. We aim at finding the minimum set of coherent states to reconstruct the POVM.
Later we will discuss how we actually choose the set of coherent states by taking into account the experimental setup. We also note that the analytical solution will not be used to reconstruct the POVM in the experiment, rather the optimization method will be used for the best possible results under the experimental errors. 

We assume a two-mode and global-phase-insensitive detector where each mode has $d$-dimensional Hilbert space. As explained in the Results, we actually have to reconstruct the POVM with larger dimensions, and then truncate it to the $d^2 \times d^2$ matrix representation. This is because we use a two-mode coherent state as the input for reconstruction. 

The general two-mode coherent state without global phase is expressed as $\ket{\alpha, \beta \ee^{i \delta}}\  (\alpha,\beta, \delta \in \mathbb{R})$, which is expanded in the photon number basis as Eq.~\eqref{eq:2modesCoherentState}.
In order to use a coherent state for the reconstruction of the POVM, we need to truncate the dimension of the coherent state so that it matches the dimension of the POVM. However, we cannot truncate the two-mode coherent state to $d^2$-dimensions.
This is because the coefficients of  $\left| m,n \right \rangle$ ($m+n=2(d-1)$) can be comparable to the coefficient of $\left| d-1,d-1 \right \rangle$. Instead we truncate the two-mode coherent state so that the total photon number is limited to $m+n \leq N$,
\begin{align}
	\ket{\alpha, \beta \ee^{i \delta}}_{N} &= 
	\exp \left[  -\frac{1}{2} \left( |\alpha|^2 +|\beta|^2 \right)
	\right]
	\sum_{m=0}^{N} \sum_{n=0}^{m} \frac{\alpha^n \beta^{(m-n)} \ee^{i (m-n) \delta}}
	{\sqrt{n! (m-n)!}} 
	\left| n,m-n \right\rangle. 
	\label{eq:Trunc2Coh}
\end{align}
The subscript of the ket, $N$, denotes the truncated vector up to the total photon number of $N$. $N$ is chosen as $N=2(d-1)$ so that the truncated Hilbert space (the dimension of $(N+2)(N+1) /2 =d(2d-1)$) contains the target Hilbert space (the dimension of $d^2$). 
Note that we will choose two amplitudes $\alpha$ and $\beta$ so that 
$~_N \langle \alpha, \beta \ee^{i \delta} | \alpha, \beta \ee^{i \delta} \rangle _{N} \simeq 1$. Henceforth we will use the truncated coherent states only, and we will omit the subscript $N$, although we may occasionally add $N$ to emphasize the truncated vector.
 
The POVM for a global phase insensitive detector with up to the total photon number of $N$ is expressed as,       
\begin{align}
\label{eq:POVM_GPI}
	\hat \Pi = \sum_{n=0}^{N} \sum_{k=0}^{n}\sum_{l=0}^{n}\pi_{k,n-k,l,n-l}\crossprojection{k,n-k}{l,n-l}.
\end{align}
This POVM has $(N+1)(N+2)(2N+3)/6$ unknowns in total. In our experiment, the dimension is $d=2$ and the total photon number should be $N=2$, thus we will need at least 14 linearly independent coherent states to reconstruct the POVM. 
As we have seen, each coherent state [Eq.~\eqref{eq:Trunc2Coh}] is associated with a triplet of parameters $(\alpha, \beta, \delta)$. A minimum set $\mathcal{S}_N$ of triplets can be expressed as
   \begin{align}
       \label{eq:MinimumSetOfCoherentStates}
        \mathcal{S}_N =&
         \left \{ \left( \alpha^{(s)}_v, \beta^{(s)}_v, \frac{2\pi}{2s+1}m \right) \right |
          s=\{0,1,\cdots, N \}, \notag \\
         &\quad 0\leq v \leq N-s, 0\leq m \leq 2s, (v, m) \in \mathbb{Z}         
      \biggr\}.
\end{align}
We will later discuss how to choose $ \left( \alpha^{(s)}_v, \beta^{(s)}_v \right)$.

Here we show how all the POVM elements can be analytically obtained with the set of coherent states $\mathcal{S}_N$. Let us first calculate the probability $p(\alpha, \beta\ee^{i \delta})$ for a two-mode coherent state input $\ket{\alpha, \beta \ee^{i \delta}}$ with the POVM $\hat \Pi$ in Eq.~\eqref{eq:POVM_GPI},
\begin{align}
	p(\alpha, \beta\ee^{i \delta}) & = \bra{\alpha, \beta \ee^{i \delta}} \hat \Pi \ket{\alpha, \beta \ee^{i \delta}} \\
	&\quad=
	\sum_{n=0}^{N} \sum_{k=0}^{n}\sum_{l=0}^{n}\pi_{k,n-k,l,n-l} \frac{ \ee ^{-\alpha ^2 -\beta ^2} \alpha ^{k+l}\beta ^{2n-k-l}}{\sqrt{k!(n-k)!l!(n-l)!}} \ee ^{i\delta (k-l)} \\
	&\quad =
	\sum_{\Delta=-N}^N \sum_{n=|\Delta|}^{N}\sum_{k=\max (0,\Delta)}^{\min (n,n+\Delta)}
	\pi_{k,n-k,k-\Delta,n+\Delta-k}C_{k,\Delta}(\alpha,\beta)  \ee ^{i\delta \Delta}, \label{eq:GeneralProb}
\end{align}
where $\Delta = k-l$, and $C_{k,\Delta}(\alpha,\beta)$ is the coefficient of $\pi_{k,n-k,k-\Delta,n+\Delta-k}$ except the phase $\ee ^{i\delta \Delta}$.
We used the relationship 
$\sum_{n=0}^{N} \sum_{k=0}^{n}\sum_{l=0}^{n} = \sum_{n=0}^{N} \sum_{\Delta=-n}^{n}\sum_{k=\max (0,\Delta)}^{\min (n,n+\Delta)}=
\sum_{\Delta=-N}^N \sum_{n=|\Delta|}^{N}\sum_{k=\max (0,\Delta)}^{\min (n,n+\Delta)}$.
Note that the number of POVM elements with $\Delta$ in Eq.~\eqref{eq:GeneralProb} is $(N-|\Delta| + 1)(N-|\Delta |+2)/2$.

By integrating this probability $p(\alpha, \beta\ee^{i \delta})$ with the phase factor $\ee^{i t \delta} (t\in \mathbb{Z})$, we can sort out the terms with $\Delta = t$ as, 
\begin{align}
&\frac{1}{2\pi}\int_{0}^{2\pi}  p(\alpha,\beta \ee^{i\delta}) \ee^{-it \delta} \dd \delta
\label{eq:Fourier_Prob}
 =  
\sum_{n=|t|}^{N}\sum_{k=\max (0,t)}^{\min (n,n+t)}
  \pi_{k,n-k,k-t,n+t-k}C_{k,t}(\alpha,\beta).
\end{align}
From the discrete set of $\delta = \frac{2\pi}{2s+1}m \in \mathcal{S}_N$, we can similarly sort out the terms with $\Delta = t$ as,
\begin{align}
	\frac{1}{(2s+1)}&\sum_{m=0}^{2s} p(\alpha,\beta \ee^{i\frac{2\pi}{2s+1}m}) \ee^{-it\frac{2\pi}{2s+1}m} 
\label{eq:Sum_Prob}
	=
	 A_t^{(s)} (\alpha,\beta) + B_t^{(s)} (\alpha,\beta) ,
\end{align}
where
\begin{align}
	 A_t^{(s)} (\alpha,\beta) &= 
	\sum_{n=|t|}^{N}\sum_{k=\max (0,t)}^{\min (n,n+t)}\pi _{k,n-k,k-t,n+t-k}C_{k,t}(\alpha,\beta), \label{eq:Ats} \\
	 B_t^{(s)} (\alpha,\beta) &= \sum_{\stackrel{\Delta = t+(2s+1)u}{(u\neq 0, |\Delta | \le N)}} \sum_{n=|\Delta |}^{N}\sum_{k=\max (0,\Delta)}^{\min (n,n+\Delta )}
	\pi _{k,n-k,k-\Delta,n+\Delta-k}C_{k,\Delta}(\alpha,\beta).
	\label{eq:Bts}
\end{align}
$ A_t^{(s)} (\alpha,\beta)$ is the desired term that is the same as Eq.~\eqref{eq:Fourier_Prob}, while $B_t^{(s)} (\alpha,\beta) $ is an additional term due to the aliasing.
Note that $B_t^{(s)} (\alpha,\beta) $ does not include any terms with the phase factor of $|\Delta| \le |t|$, if $|t| \le s$.

We will then explain the sequential procedure to calculate the POVM elements using Eq.~\eqref{eq:Sum_Prob}. We assume that we have a set of probabilities $p(\alpha,\beta \ee^{i \delta})$ for $(\alpha,\beta, \delta) \in \mathcal{S}_N$.
First we use data for the input coherent states of
$\Bigl \{
          \left( \alpha^{(N)}_0, \beta^{(N)}_0, \frac{2\pi}{2N+1}m \right) 
          \Bigl | 
          0 \leq m \leq 2N, m \in \mathbb{Z}         
 \Bigr \}$.
Let $s=t=N$ in Eqs.~(\ref{eq:Ats}) and (\ref{eq:Bts}), which give us,
\begin{align}
A_N^{(N)} (\alpha_0^{(N)},\beta_0^{(N)}) &=\pi _{N,0,0,N}C_{N,N}(\alpha_0^{(N)},\beta_0^{(N)}), \\
B_N^{(N)} (\alpha_0^{(N)},\beta_0^{(N)}) &=0.
\end{align}
Since the left-hand side of Eq.~\eqref{eq:Sum_Prob} is computable from the given probabilities, we can calculate the POVM element $\pi _{N,0,0,N}$ (or the POVM element for $\Delta =N$ in the expression of Eq.~\eqref{eq:GeneralProb}).
Next, we calculate $A_{N-1}^{(N)}(\alpha_0^{(N)},\beta_0^{(N)})$ and $B_{N-1}^{(N)}(\alpha_0^{(N)},\beta_0^{(N)})$ (i.e., $s=N$, $t=N-1$) using the same data set.
In this case, $A_{N-1}^{(N)}(\alpha_0^{(N)},\beta_0^{(N)})$ contains three POVM elements for $\Delta =N-1$, i.e.,
$\pi _{N-1,0,0,N-1}$, $\pi _{N-1,1,0,N}$, and $\pi _{N,0,1,N-1}$, while  $B_{N-1}^{(N)}(\alpha_0^{(N)},\beta_0^{(N)})$ is still zero. Thus we have an equation with three unknowns. To solve this equation, 
we use additional data for the input coherent states of 
$\Bigl \{
          \left( \alpha^{(N-1)}_v, \beta^{(N-1)}_v, \frac{2\pi}{2N-1}m \right) 
          \Bigl | v=\{0,1 \}, 
          0 \leq m \leq 2(N-1), m \in \mathbb{Z}         
 \Bigr \}$.
Letting $s=N-1$ and $t=N-1$ in Eq.~\eqref{eq:Sum_Prob} with these data, we obtain two additional equations for $v=0,1$. 
In this case, $A_{N-1}^{(N-1)} (\alpha_v^{(N-1)},\beta_v^{(N-1)})$ contains the same three POVM elements for $\Delta =N-1$, while 
 $B_{N-1}^{(N-1)} (\alpha_v^{(N-1)},\beta_v^{(N-1)})$ contains $\pi _{0,N,N,0}$.
Since $\pi _{0,N,N,0} = \pi _{N,0,0,N}^*$ is already derived, we can solve the simultaneous equations regarding the POVM elements for $\Delta =N-1$ as long as these equations are linearly independent.

In the same manner, we can sequentially calculate the POVM elements for $\Delta =N -l$.
When we set $t=N-l$ in Eq.~\eqref{eq:Sum_Prob} with the set of inputs 
$\Bigl \{
          \left( \alpha^{(s)}_v, \beta^{(s)}_v, \frac{2\pi}{2N-1}m \right) 
          \Bigl | 
          s=\{ N-l, N-l+1, \cdots, N \},~ 0\leq v \leq N-s, ~
          0 \leq m \leq 2s,~ \{v,m \} \in \mathbb{Z}         
 \Bigr \}$,
we will obtain $(l+1)(l+2)/2$ equations with the unknown POVM elements for $\Delta =N -l$ and the known POVM elements for $|\Delta|>N-l $. 
Since the number of the unknown POVM elements for $\Delta =N -l$ is $(N-|\Delta| + 1)(N-|\Delta |+2)/2 = (l+1)(l+2)/2 $, we can solve those simultaneous equations as long as the equations are linearly independent.

Thus, we have proven that the set of coherent states in Eq.~\eqref{eq:MinimumSetOfCoherentStates} is enough to calculate all the POVM elements. 
Since the degrees of freedom of the POVM is the same as the number of input coherent states, the set of coherent states in Eq.~\eqref{eq:MinimumSetOfCoherentStates} is a minimum set of input states. 
The additional constraints are that we have to choose $( \alpha^{(s)}_v, \beta^{(s)}_v )$ so that the simultaneous equations derived in the procedure are linearly independent, and the truncation is adequately achieved, i.e., 
$~_N \langle 
\alpha^{(s)}_v, \beta^{(s)}_v \ee^{i \delta} 
| 
\alpha^{(s)}_v, \beta^{(s)}_v \ee^{i \delta} 
 \rangle_N \simeq 1$.

\section{Preparation of coherent state inputs}
We will explain how to prepare coherent states in the experiment. The minimum number of coherent states is 14 as explained before.
We will actually prepare a larger number of coherent states by taking into account experimental implementations.

\figureref{ExperimentalSetup}(c)
 shows the schematic to prepare a coherent state in the polarization two-modes, $\ket{\alpha, \beta}\equiv \ket{\alpha}_H\ket{\beta}_V$.
The optical power of the horizontally polarized beam after the polarization beam splitter (PBS$_1$) is denoted as ${P}$, i.e., the state vector is expressed as $\ket{\sqrt{P},0}$.
The state after the quarter wave plate (QWP$_2$) and the half wave plate (HWP$_2$) with the angles $\theta_Q$ and $\theta_H$ is,
\begin{align}
	\ket{\alpha,\beta} \equiv
	\hat U _\mathrm{HWP}(\theta_H ) 
	\hat U _\mathrm{QWP}(\theta_Q ) \ket{\sqrt{P},0},
\end{align}
where $\hat U _\mathrm{QWP}$ and $\hat U _\mathrm{HWP}$ are unitary operators for QWP and HWP. 
We use the Jones matrix method \cite{Yariv07} to calculate this, 
\begin{align}
	&
	\begin{pmatrix}
	\alpha \\ \beta
	\end{pmatrix}
	=	
	U_\mathrm{HWP}(\theta_H ) 
	U_\mathrm{QWP}(\theta_Q ) 
	\begin{pmatrix}
	\sqrt{P} \\ 0
	\end{pmatrix},
\end{align}
where
\begin{align}
	U_\mathrm{HWP}(\theta_H ) &=
		\begin{pmatrix}
		\cos 2\theta_H & \sin 2\theta _H \\
		\sin 2\theta _H & -\cos 2\theta_H
	\end{pmatrix},\\
	U_\mathrm{QWP}(\theta_Q ) &=
			\frac{1}{\sqrt{2}}
	\begin{pmatrix}
		1-i\cos 2\theta_Q & -i\sin 2\theta _Q \\
		-i\sin 2\theta _Q & 1+i \cos 2\theta_Q
	\end{pmatrix}.
\end{align}
Let us calculate the absolute values of the output amplitudes ($\alpha$ and $\beta$), and the relative phase angle $\delta$,
\begin{align}
	|\alpha| & =\sqrt{\frac{P}{2}\left(1+\cos 2\theta_Q \cos 2(2\theta_H-\theta_Q )\right)}, \\
	|\beta| & = \sqrt{\frac{P}{2}\left(1-\cos 2\theta_Q \cos 2(2\theta_H-\theta_Q )\right)}, \\
	\tan\delta &= \dfrac{\tan 2\theta_Q}{\sin 2(2\theta_H-\theta_Q )},
\end{align}
where $\alpha = |\alpha | \ee ^{i \delta_\alpha}$, $\beta = |\beta | \ee ^{i \delta_\beta}$, and $\delta = \delta_\beta -\delta_\alpha$.
Thus, we can prepare an arbitrary two-mode coherent state $\ket{|\alpha|,|\beta|\ee^{i\delta}}$, by adjusting the optical power $P$ and the wave plate angles $\theta_Q$ and $\theta_H$.

In order to reconstruct the detector POVMs with $N=2$, 
we prepare 19 sets of coherent states as listed in Table \ref{tb:WavePlateSet}. These states are represented by the parameters in the set as shown in Eq.~\eqref{eq:set}. 
Note that 
$~_N  \langle \sqrt{0.2},0  
| 
\sqrt{0.2}, 0  \rangle_N =0.999
$, and we decided to employ slightly larger number of coherent states to simplify the rotating angles of two wave plates.

\begin{table}[H]
\centering
\caption{Input states preparation in the experiment. $\theta _Q$ and $\theta _H$ are the rotating angles of QWP and HWP, respectively.
	The units of powers $P$, $|\alpha |^2$, and $|\beta |^2$ are the photon numbers per wave packet.}
\label{tb:WavePlateSet}
\begin{tabular}{rrrrrr}
\arrayrulecolor{black}\hline
\multicolumn{1}{c}{$P$} & \multicolumn{1}{c}{$\theta_Q$[deg]} & \multicolumn{1}{c}{$\theta_H$[deg]} & \multicolumn{1}{c}{$\quad |\alpha| \quad$} & \multicolumn{1}{c}{$\quad|\beta|\quad$} & \multicolumn{1}{c}{$\delta$[deg]} \\ \hline
0.20                  & $-22.5$                           & $-33.75$                          & 0.316                   & 0.316                   & $-135$                           \\
0.20                  & $-45$                             & $-22.5$                           & 0.316                   & 0.316                   &$ -90$                            \\
0.20                  & $-22.5$                           & 11.25                           & 0.316                   & 0.316                   &$ -45$                            \\
0.20                  & 0                               & 22.5                            & 0.316                   & 0.316                   & 0                              \\
0.20                  & 22.5                            & 33.75                           & 0.316                   & 0.316                   & 45                             \\
0.20                  & 45                              & 22.5                            & 0.316                   & 0.316                   & 90                             \\
0.20                  & 22.5                            & $-11.25$                          & 0.316                   & 0.316                   & 135                            \\
0.20                  & 0                               & $-22.5$                           & 0.316                   & 0.316                   & 180                            \\
0.20                  & 0                               & 0                               & 0.447                   & 0                       & \multicolumn{1}{c}{-}          \\
0.20                  & 0                               & 45                              & 0                       & 0.447                   & \multicolumn{1}{c}{-}          \\
0.05                  & $-30$                             & $-15$                             & 0.194                   & 0.112                   &$ -90$                            \\
0.05                  & 0                               & 15                              & 0.194                   & 0.112                   & 0                              \\
0.05                  & 30                              & 15                              & 0.194                   & 0.112                   & 90                             \\
0.05                  & 0                               & $-15$                             & 0.194                   & 0.112                   & 180                            \\
0.05                  & $-30$                             & 30                              & 0.112                   & 0.194                   & $-90$                            \\
0.05                  & 0                               & 30                              & 0.112                   & 0.194                   & 0                              \\
0.05                  & 30                              & $-30$                             & 0.112                   & 0.194                   & 90                             \\
0.05                  & 0                               &$-30$                             & 0.112                   & 0.194                   & 180                            \\
0                     & \multicolumn{1}{c}{-}           & \multicolumn{1}{c}{-}           & 0                       & 0                       & \multicolumn{1}{c}{-}          \\ \hline
\end{tabular}
\end{table}

\section*{Funding}
This work was supported by the Australian Research Council Centre of Excellence for Quantum Computation and Communication Technology (Project No. CE170100012).

\section*{Acknowledgments}
The authors would like to thank Tim Ralph for his valuable comments, and Thomas d'Ews Thomson and Aleksandar Davidovic for their technical supports.
S.Y. acknowledges support from JSPS Overseas Research Fellowships of Japan.


	
\end{document}